\documentclass{article}
\usepackage{epsfig}
\begin{document}
\newcommand{\dd} {\mbox{d\raisebox{0.75ex}{\hspace*{-0.32em}-}\hspace*{-0.02em}}}
\newcommand{\DD} {\mbox{D\raisebox{0.30ex}{\hspace*{-0.75em}-}\hspace*{ 0.42em}}}
\begin{center}

{ \bf \Large A Search For Solar Hadronic Axions Using $^{83}Kr$} \\
\vspace{1cm}
 { \em K. Jakov\v ci\' c, Z. Kre\v cak\footnote{Corresponding author. Fax: +385-1-468-0239.
 Email address: krecak@rudjer.irb.hr (Z. Kre\v cak)}, M. Kr\v cmar and A. Ljubi\v ci\' c\\
 Department of Experimental Physics, Ru\dd er Bo\v skovi\' c Institute, \\
 P.O. Box 180, Bijeni\v cka cesta 54, 10002 Zagreb, Croatia }
\vspace{1cm}

 \abstract{We introduce a new experimental method for solar hadronic axions search.
 It is suggested that these axions are created in the Sun during M1 transition between the
 first thermally excited level at 9.4 keV and the ground state in $^{83}Kr$. Our method is
 based on axion detection via resonant absorption process by the same nucleus in the
 laboratory. We use proportional gas counter filled with krypton to detect signals for axions.
 With this setup, target and detector are the same which increases the efficiency of the
 experiment. At present, an upper limit on hadronic axion mass of 5.5 keV at the
 $95\%$ confidence level is obtained. }

\end{center}

\section{Introduction}

Axions, neutral spin-zero pseudoscalar particles, arise from
spontaneous breaking of the Peccei-Quinn (PQ) chiral symmetry, the
latter being introduced to resolve the strong CP problem. In
general, they interact with leptons, photons and hadrons. Axion
mass can be interpreted as a mixing of the axion field with pions,
and is related to the PQ symmetry breaking scale $f_{a}$ by
$m_{a}f_{a}\approx m_{\pi}f_{\pi}$ where $f_{\pi}$ denotes pion
decay constant. The original suggestion ($m_{a}\approx 100$ keV)
has not been experimentally confirmed and DFSZ/GUT and KSVZ models
of invisible axions have been developed. KSVZ axion coupling to
leptons is suppressed. Astrophysical and cosmological
considerations predict axion mass window $10^{-5}eV \leq m_{KSVZ}
\leq 10^{-2}eV$ relating to cold dark matter. Hadronic axion is a
class of KSVZ axions which does not couple to photon. Its mass
window $10\; eV < m_{a} < 40\; eV$ relates to hot dark matter.
Moriyama was the first to propose the production of monochromatic
axions in the solar interior during M1 transitions between first
thermally excited 14.4  keV and ground state in $^{57}Fe$
\cite{Moriyama1995}. When resonant conditions are fulfilled, the
emitted axion could excite same nucleus at long distance. This
event could be detected through the subsequent emission of a
photon or conversion electron. In our previous experiments we
searched for axion-produced M1 electromagnetic transitions in
$^{57}Fe$ \cite{Krcmar1998} and $^{7}Li$ \cite{Krcmar2001} using
Si(Li) and HPGe detectors, respectively. Our experiments yield
upper limits on the hadronic axion mass of 745 eV and 32 keV,
respectively.

\section{Experimental method}
In this experiment we were looking for the M1 transitions from the
first excited 9.4 keV to the ground state in $^{83}Kr$ as possible
signals for the resonant absorption of axions emitted from the Sun
by the same nucleus. Due to the 1.35 keV thermal motion of
$^{83}Kr$ nuclei in the Sun core, the extremely narrow natural
line width of emitted axions  is Doppler broadened to $\cong 2.93$
eV. Due to the emission and absorption nuclear recoil, the axion
energy shift is $-2\cdot 0.000572$ eV. Gravitational red shift of
axions moving from Sun to Earth is  $-0.101$ eV. Because recoil
and gravitational red shifts are much smaller than Doppler
broadening in the Sun we conclude that conditions for resonant
absorption of axions are fulfilled. The stable isotope of krypton,
$^{83}Kr$ (with natural abundance $11.5 \%$) is reasonably well
abundant in the Sun; from the experimentally determined relative
abundance of krypton to hydrogen in meteorites $\cong 1.7\cdot
10^{-9}$, the solar abundance of $^{83}Kr$ by mass fraction is
estimated to be $\cong 2\cdot 10^{-10}$. For detection of 9.4 keV
photons and lower energy conversion electrons we used proportional
counter filled with natural krypton. In this arrangement, target
and detector are the same which increases the sensitivity of the
experiment.\\

The excitation rate $R$ of the 9.4 keV state in $^{83}Kr$ on Earth
was estimated using the KSVZ axion model \cite{Kaplan1985,
Srednicki1985}.  Nuclear data about M1 transition between ground
9/2+ and first excited 7/2+ state ($T_{1/2} = 154.4$ ns,
$\alpha=17.09$) in $^{83}Kr$ were used from \cite{Wu2001}. In
one-particle approximation the transition is considered as neutron
transition which gives nuclear-structure dependent terms $\eta =
0.5$ and $\beta = -1$. With these approximations we obtained
\begin{equation}
R\left[ g^{-1}\; day^{-1} \right] = 7.417\cdot 10^{-12}\left( \frac{k_{a}}{k_{\gamma}}\right) ^{6}
\cdot \left( \frac{m_{a}}{1\; eV}\right) ^{4}
\end{equation}

 where $k_{a}$ and $k_{\gamma}$ denote impulses of axions and photons,
respectively.

\section{Experimental setup}
We have used a small proportional gas counter and self-made preamplifier designed
for detection of low energy photons and electrons with energies from about 5 keV to
about 100 keV. Its cylindrical brass chamber, with effective inner diameter and length
 of $45$ mm and $153$ mm, respectively, could be filled by various gases, biased up
 to $3$ kV and could operate at stable pressures from $0.0000001$ bar up to $4$ bars.
 Its steel anode with diameter of 50 $\mu m$ is connected via its MHV connector to the
 preamplifier which is supplied with power, signal and test connectors.
 The chamber is equipped with three apertures for good quality valve,
 adequate pressure gauge and photon collimator. The valve keeps gas in the chamber
 at steady pressure up to 4 bars for at least a year. The aperture with diameter
 of 1 mm for photon collimator is placed perpendicular to the chamber axis and
 in the middle along the chamber. Thickness of the beryllium window is 5 $\mu m$.\\

The system was tuned for detection of photons with energies from 5 to 25 keV by
applying the anode  bias voltage of $+1600$ V. It was calibrated by the use of $^{55}Fe$
 and $^{109}Cd$ X-ray emitters. Energy resolution of proportional counter
 at 9.4 keV was 4.7 keV.

\section{Results and discussion}
Expected number of counts $N_{c}$ in our detector, due to the absorption of axions
 and subsequent de-excitation of the 9.4 keV state, is
\begin{equation}
N_{c}=R\left[ g^{-1}\; day^{-1} \right]\cdot M[g]\cdot \Delta T[day]\cdot \varepsilon
\end{equation}

where $M$ is mass of $^{83}Kr$ atoms in the counter, $\Delta T$ is time of data collection,
 and $\varepsilon$ is detection efficiency. In our experiment we kept the pressure inside
 the counter at 2 bar and therefore $M=0.193$ grams. From the tables of electron and photon
 stopping powers and taking into account small counting rate we estimated
 $\varepsilon \approx 0.99$. \\

Data were collected over period of $23.5$ days. We did not observe any evidence of
statistically significant axion events. Therefore an upper limit on hadronic axion mass
was determined from the expression
\begin{equation} \label{izraz3}
 \frac{m_{a}}{1\; eV}\cdot \left( \frac{k_{a}}{k_{\gamma}}\right) ^{3/2}<
 \left( \frac{k^{'}\cdot \sqrt{2\cdot N_{b}}}{7.417\cdot 10^{-12}
 \cdot M[g]\cdot \Delta T[day]\cdot \varepsilon }\right)^{1/4}
\end{equation}
Here $N_{b}$ is number of background events in the relevant energy interval.
At the $95 \%$ CL the recorded number of background events was  $N_{b} \cong 1.73\cdot 10^{8}$
 cts. Using Eq (\ref{izraz3}) an upper limit on hadronic axion mass of 5.5 keV is obtained.

\section{Conclusion}
We have performed an experimental search for solar hadronic axions using a new experimental
 method in which the target and the detector were the same.
 In our further research we will try to improve background suppression,
 energy resolution and detector volume. This could enable us to search for
 hadronic axion masses in the region $<100$ eV.\\

\paragraph{Acknowledgement}
\- \\
We thank the Ministry of Science and Technology of Croatia for financial support (contract No. 0098011). We thank  K. Kova\v cevi\' c
 for electronic service.



\begin{thebibliography}{NefE890}

\bibitem{Moriyama1995} Moriyama S., 1995. Proposal to search for
a monochromatic component of solar axions using $^{57}Fe$. Phys.
Rev. Lett. 78. 3222-325.

\bibitem{Krcmar1998} Kr\v cmar M., Kre\v cak Z.,  Stip\v cevi\' c M., Ljubi\v ci\' c A.,
 and Bradley D.A., 1998. Search for solar axions using $^{57}Fe$. Phys.
Lett. B 442. 38-42.

\bibitem{Krcmar2001}Kr\v cmar M.,  Kre\v cak Z., Ljubi\v ci\' c A.,
Stip\v cevi\' c M., and  Bradley D.A., 2001. Search for solar
axions using  $^{7}Li$. Phys. Rev. D 44. 115016-1/4

\bibitem{Kaplan1985} Kaplan D.B., 1985. Opening the axion window. Nucl.
Phys. B 260. 215-226.

\bibitem{Srednicki1985} Srednicki M., 1985. Axion couplings to matter.
Nucl. Phys. B 260. 689-700.

\bibitem{Wu2001} Wu S.-C., 2001. Nuclear data sheets / Revised
evaluations for A = 83. Nucl. Data Sheets 92. 925-942.

\end{thebibliography}
\end{document}